\documentclass[manuscript,screen]{acmart}
\usepackage[capitalise]{cleveref}
\usepackage{tabularx}

\AtBeginDocument{%
  }

\setcopyright{acmlicensed}
\copyrightyear{2025}
\acmYear{2025}
\acmDOI{XXXXXXX.XXXXXXX}
\acmISBN{978-1-4503-XXXX-X/2018/06}

\begin{document}

\title[The Screen4Care Metadata Repository]{S4CMDR: a metadata repository for electronic health records}


\author{Jiawei Zhao}
\email{jiz@sdu.dk}
\orcid{}
\author{Md Shamim Ahmed}
\email{shamim@imada.sdu.dk}
\orcid{}
\author{Nicolai Dinh Khang Truong}
\email{nitru@imada.sdu.dk}
\orcid{}
\affiliation{%
  \institution{University of Southern Denmark, Department for Mathematics and Computer Science}
  \city{Odense}
  \country{Denmark}
}

\author{Verena Schuster}
\email{VSchuster@sba-research.org}
\orcid{}
\author{Rudolf Mayer}
\email{RMayer@sba-research.org}
\orcid{}
\affiliation{%
  \institution{SBA Research}
  \city{Vienna}
  \country{Austria}
}

\author{Richard R\"ottger}
\email{roettger@imada.sdu.dk}
\orcid{}
\affiliation{%
  \institution{University of Southern Denmark, Department for Mathematics and Computer Science}
  \city{Odense}
  \country{Denmark}
}

\renewcommand{\shortauthors}{Zhao et al.}

\begin{abstract}
\textbf{Background:} Electronic health records (EHRs) enable machine learning for diagnosis, prognosis, and clinical decision support. However, EHR standards vary by country and hospital, making records often incompatible. This limits large-scale and cross-clinical machine learning. To address such complexity, a metadata repository cataloguing available data elements, their value domains, and their compatibility is an essential tool. This allows researchers to leverage relevant data for tasks such as identifying undiagnosed rare disease patients.

\textbf{Results:} Within the Screen4Care project, we developed S4CMDR, an open-source metadata repository built on ISO 11179-3, based on a middle-out metadata standardisation approach. It automates cataloguing to reduce errors and enable the discovery of compatible feature sets across data registries. S4CMDR supports on-premise Linux deployment and cloud hosting, with state-of-the-art user authentication and an accessible interface.

\textbf{Conclusions:} S4CMDR is a clinical metadata repository registering and discovering compatible EHR records. Novel contributions include a microservice architecture, a middle-out standardisation approach, and a user-friendly interface for error-free data registration and visualisation of metadata compatibility. We validate S4CMDR's case studies involving rare disease patients. We invite clinical data holders to populate S4CMDR using their metadata to validate the generalisability  and support further development.
\end{abstract}



\received{xxx}
\received[revised]{xxx}
\received[accepted]{xxx}

\maketitle

\section{Introduction \& Background}\label{background}
\subsection{Electronic Health Records for Rare Diseases} \label{EHR_RD_intro}
Electronic health records (EHRs) can be defined as "data pertaining to an individual’s health record available in a suitable electronic form that makes it machine-interpretable" \cite{IEEE_EHR_definition}. With the growing adoption of EHRs in clinical institutions and the increasing capabilities of machine learning (ML) algorithms, these machine-learning techniques are increasingly leveraged to support diagnosis, prognosis, and clinical decision-making. However, the success of every ML endeavor is heavily reliant on high-quality and comprehensive data, allowing the identification of the subtle patterns in the data critical for the correct decision of the model. This challenge is particularly aggravated when studying rare diseases (RDs), where the number of available cases is necessarily significantly smaller compared to common diseases.

The definition of RDs varies globally. For instance, the EU defines it as a life-threatening or chronically debilitating disease of predominantly genetic origin that occurs in less than 5 per 10,000 people in the community \cite{EU_RD_summary}. The number of distinct RDs is estimated to be between 6,000 and 8,000 today, ranging from relatively common RDs to ultra-rare diseases affecting only a few patients worldwide \cite{EU_RD_summary}. Therefore, it can be easily concluded that RDs typically have a scarce and virtually unfathomable nature. Nevertheless, there are already successful approaches to leverage ML for RD. Examples include detecting lipodystrophy has been made possible using a combination of unsupervised feature selection, unsupervised clustering, and supervised ensemble learning \cite{Colbaugh2018detectingLipodystrophy}; identifying misdiagnosed acute hepatic porphyria using a support vector machine \cite{Cohen2020detectingAHP}; or identifying potential wild-type transthyretin amyloid cardiomyopathy based in medical claims data in its EHR using a support vector machine \cite{Rudolph2021_ML_RD}.

\subsection{Challenges in learning on Electronic Health Records} \label{ML_on_EHR}
Despite these few positive examples of success, researchers often encounter significant obstacles while developing ML algorithms to diagnose RDs from EHRs. Combining EHRs is often necessary, but finding compatible data sources is difficult. This is due to heterogeneous medical ontology standards and data structures adopted by different EHR data sources. Commonly used medical ontology standards, e.g., the Systematized Nomenclature of Medicine Clinical Terms (SNOMED-CT) \cite{snomedct}, the International Classification of Diseases (ICD) in the ICD-10 \cite{icd-10} or ICD-11 \cite{icd-11} versions, the Logical Observation Identifiers Names and Codes (LOINC) \cite{loinc}, the Orphanet Rare Disease Ontology (ORDO) \cite{ordo}, NCI Thesaurus \cite{ncit}, and the Human Phenotype Ontology (HPO) \cite{Gargano2024_HPO} each classify clinical data differently, contain information in different granularities, and cover different aspects of the necessary clinical data. The first required step is thus the discovery of data registries holding compatible data to form a cohort of sufficient size for subsequent ML. 

Another impediment is data privacy. In recent decades, data protection regulations, e.g., the Health Insurance Portability and Accountability Act \cite{HIPAA}, the California Privacy Rights Act \cite{CPRA}, and the General Data Protection Regulation \cite{GDPR} in the EU control the sharing of sensitive EHR data. Failing to comply can result in legal issues.
Certain strategies, such as differential privacy~\cite{dwork_differential_2006}, federated learning, and secure multi-party computation, help address these issues. However, the fundamental challenge remains: data must be discovered and standardized in a compatible format on a technical level and a semantic level.

\subsection{Necessity of a clinical Metadata Repository} \label{MDR_necessity}
To enable successful collaboration on ML projects involving EHRs, particularly for RDs, the discovery of suitable and compatible data registries is imperative. As properly curated metadata are instrumental in facilitating data integration across heterogeneous databases for supporting researchers in their investigations \cite{Dugas2016}, we thus propose a new metadata repository (MDR) architecture that contains metadata, i.e., fine-grained descriptive information on the data without containing the data itself, such as its characteristics, varieties, and the standardisation of different data sources. To populate and seamlessly manage an MDR, a metadata management system is required. The system stores existing metadata and supports researchers to ingest non-redundant metadata from new data registries into the repository, adhering closely to existing ontologies to ensure rapid usability and adaptability.

\subsection{ISO 11179-3 Metadata Repository Standard} \label{ISO11179}
\begin{figure}[h]
\centering
\includegraphics[width=0.9\textwidth]{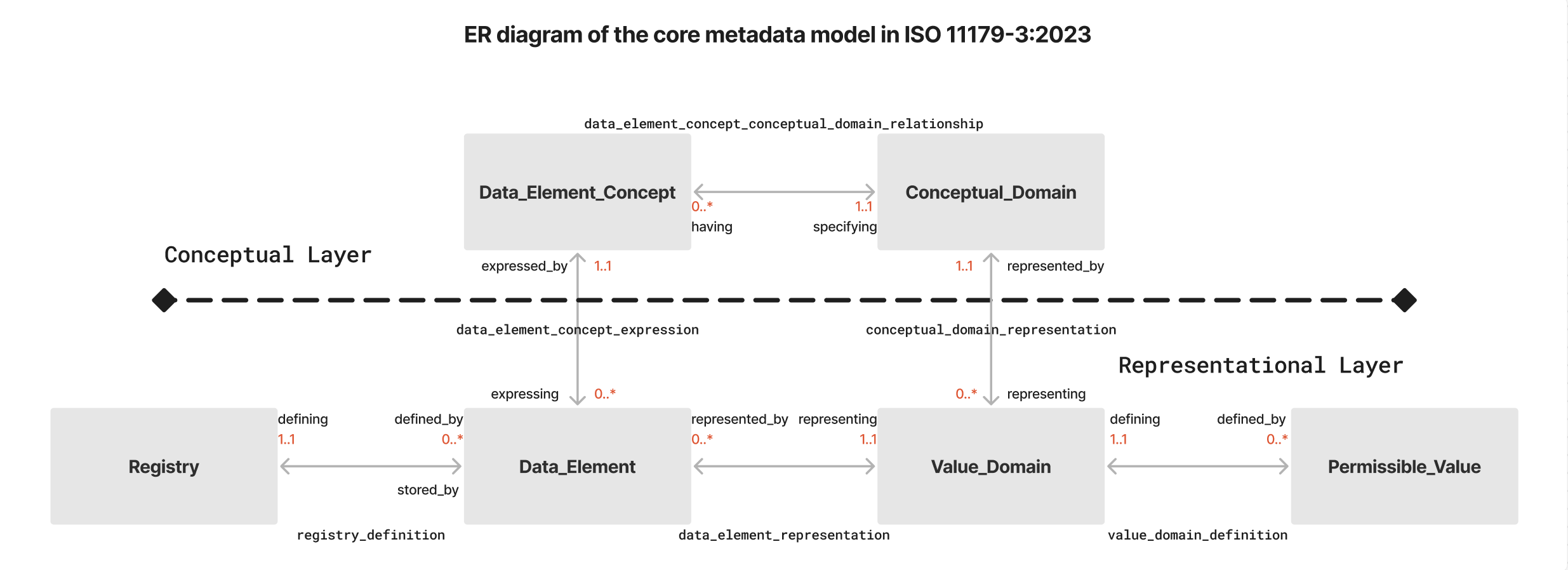}
\caption{Six crucial metadata classes are included in this figure. The wording of the figure is adopted for this manuscript.}
\label{fig.iso11179_model}
\end{figure}

Given the complexity of metadata standardisation, which later is described in Section \ref{md_standardisation_intro}, and the requirements of developing a user-friendly MDR, the MDR data structure must be modelled appropriately. To address these challenges, the ISO 11179-3 standard has been jointly developed by the International Organisation for Standardisation and the International Electrotechnical Commission \cite{ISO-11179} to provide metadata classes, their attributes, and relationships. The standard introduces a core model that formally represents metadata, divided into two layers. One layer is a "representational" one consisting of \textit{Data\_Element} and \textit{Value\_Domain}. Another layer is a "conceptual" one consisting of \textit{Data\_Element\_Concept} and \textit{Conceptual\_Domain}. \cref{fig.iso11179_model} shows an Entity-Relationship (ER) diagram of the ISO 11179-3 core metadata model. The ISO 11179-3 is a general standard and is not necessarily specific to clinical data or EHRs. The data structure developed in S4CMDR implements a modified version of these classes to represent the stored metadata.

Regarding the clinical interpretation of the core metadata model shown in \cref{fig.iso11179_model}, \textit{Data\_Element} is the storage path in a clinical data registry (for example, diagnosis coded by SNOMED CT or laboratory test value coded by LOINC). A \textit{Data\_Element\_Concept}, each expressed by a set of \textit{Data\_Elements} in various data registries, represents one abstract feature in an EHR, usually coded by a medical ontology class. Proper annotation of \textit{Data\_Element\_Concepts} is essential for feature selection and virtual cohort creation. A \textit{Conceptual\_Domain} represents an abstract aggregation of \textit{Data\_Element\_Concepts} representing semantically related medical ontology classes or other clinical features. A \textit{Value\_Domain} is a metadata class annotated with attributes such as datatype and format, and defines a set of \textit{Permissible\_Values} that determine the permitted values of \textit{Data\_Elements} and are essential to indicate the compatibility between the data registries.

\subsection{Metadata Standardisation Approaches} \label{md_standardisation_intro}
Three metadata standardisation approaches are \textit{top-down}, \textit{bottom-up}, and \textit{middle-out} \cite{Kim2019_Clinical_MD_ontology}.
The \textit{top-down} approach is centralized, allowing authorities, e.g., from SNOMED CT and LOINC, to define metadata items.
However, the \textit{top-down} approach is tedious to implement in real-world healthcare scenarios \cite{Tennison2002_APECKS}.
The decentralised \textit{bottom-up} approach allows users to construct their clinical concepts (e.g., by combining \textit{injury}, \textit{back}, \textit{head}, \textit{neck} to describe patients' injuries)\cite{Roberts2015_AutomaticExtraction}. However, this approach is impractical due to the challenge and complexity of encoding clinical concepts uniformly.
The \textit{middle-out} approach offers a balance by focusing on the semantic interoperability of clinical \textit{Data\_Elements} in EHRs \cite{Kim2019_Clinical_MD_ontology}, based on the ISO 11179-3 standard, and by improving "common data elements" in MDRs.
We opted for the \textit{middle-out} approach as it pragmatically defines the metadata items needed at a highly practical level for immediate real-world use \cite{Kim2019_Clinical_MD_ontology}.

\subsection{Our Contribution} \label{S4CMDR_intro}
In this manuscript, we introduce a novel, scalable, and demand-tailored metadata repository architecture for electronic health records, called S4CMDR. It was developed within the framework of the EU-funded "Innovative Medicine Initiative" (IMI) project Screen4Care~\cite{Ferlini2023_S4C}, which seeks to shorten the time to diagnosis of RD patients. Our implementation is based on industry-standard technologies and provides capabilities, guided entry of data concepts, discovery of data registries, checks for data compatibility, secure user management, and user-friendly visualisations.

Our concrete contributions are as follows:
\begin{itemize}
    \item Analysis of the ISO 11179-3 core metadata model and its shortcomings regarding the effective modelling of the commonly-used medical ontology standards, and our adaptive modifications to the core metadata standard
    \item Summary of the existing ISO 11179-based MDRs and the analysis of their absent features
    \item Development of a comprehensive MDR architecture based on state-of-the-art technologies
    \item Secure user management and the ability to connect to other services through Keycloak
    \item Connection to existing medical ontologies in the MDR or BioPortal, guiding the registration of new data registries
    \item Visualisation of the compatibility of data registries and discovery of suitable data registries for ML
    \item  Validation of the ingestion of metadata using a demonstration dataset with a focus on RDs
    \item The S4CMDR is available in a public git repository such that researchers can host their own on-premise repository, and contribute to the development of S4CMDR
\end{itemize}

Following the \nameref{background} section, the remainder of this manuscript is organized as follows: \nameref{RelatedWork} provides an overview of MDRs, highlighting their strengths and weaknesses and the gaps that motivate this work. The \nameref{impl} section illustrates the original ISO 11179-3 core metadata standard and its curated version adapted to the common medical ontology standards, and provides details of the implementation of S4CMDR, a comprehensive explanation of the modifications applied to the core data model of the S4CMDR compared to the original ISO 11179-3 standard and the implemented microservice architecture. The \nameref{results} section introduces the functionalities and user interface (UI) in S4CMDR, emphasizing its unique features and advantages over the existing published MDRs. The \nameref{discussion} section addresses the limitations and potential future improvements of S4CMDR, followed by our conclusions.
\section{Related Work}\label{RelatedWork}
Dedicated metadata repositories for EHRs have been developed before.
We want to highlight the necessary requirements and compare existing solutions here:
\begin{itemize}
\item The findable, accessible, interoperable, and reusable (FAIR) principles are crucial for effective metadata management and integration \cite{Wilkinson2016_FAIR}. 
MDRs must prioritize adhering to FAIR principles. 
In particular, the metadata items must be easily findable and interoperable across data sites.
\item Careful alignment across heterogeneous medical ontologies (e.g., SNOMED CT \cite{snomedct}, NCI Thesaurus \cite{ncit}, LOINC \cite{loinc}, and ICD-10 \cite{icd-10}) is necessary to minimize redundancy and errors in the MDR's vocabulary.
\item Metadata items must be validated and curated for medical/clinical relevance. These could be achieved by, for example, utilizing BioPortal APIs \cite{bioportal-api}, allowing medical experts to assess and create relevant inputs.
\item A community-driven open-source MDR is important because it enables users to update and maintain the latest metadata items.
\end{itemize}

Although several ISO 11179-based MDR architectures have been introduced, e.g., caDSR \cite{caDSR}, Samply.MDR \cite{SamplyMDR}, QL$^4$MDR \cite{ql4mdr} and Pragmatic MDR \cite{pragmaticMDR}, none of them were developed with the aforementioned requirements in mind nor do they support a curated core metadata model suitable for conceptual and representational adaptation to the common medical ontology standards.  Moreover, these MDRs have primarily focused on the representational layer of the ISO 11179 core metadata standard (See \Cref{fig.iso11179_model}). 

We provide a structured summary of the aforementioned ISO 11179-based MDRs published in \cref{tbl.existing_iso11179_mdr}. For detailed explanations of the ISO 11179-based core metadata model adopted in S4CMDR, please refer to \Cref{impl.core_data_model} and \Cref{fig.s4c_mdr_data_model}. 

\begin{table}
\centering 
\caption{Summary of the existing published ISO 11179-based MDRs.} 
\begin{tabularx}{\textwidth}{l X}
\toprule
Name & Implementation Details \\
\midrule
caDSR \cite{caDSR} & Emphasis of the data model: \textbf{conceptual and representational} \\
      & Approach for metadata standardisation: \textbf{top-down} \\
      & Year of journal publication: \textbf{2003} \\
      & Open-source code available: \textbf{yes} \\ 
      & Link of open-source code: \url{https://github.com/orgs/NCIP/repositories?q=cadsr} \\ 
    \midrule
Samply.MDR \cite{SamplyMDR} & Emphasis of the data model: \textbf{representational} \\
           & Approach for metadata standardisation: \textbf{top-down} \\
           & Year of journal publication: \textbf{2018} \\
           & Open-source code available: \textbf{yes} \\ 
           & Link of open-source code: \url{https://bitbucket.org/medicalinformatics/workspace/repositories/?search=samply} \\ 
    \midrule
QL$^4$MDR \cite{ql4mdr} & Emphasis of the data model: \textbf{representational} \\
       & Approach for metadata standardisation: \textbf{middle-out} \\
       & Year of journal publication: \textbf{2019} \\
       & Open-source code available: \textbf{yes} \\ 
       & Link of open-source code: \url{https://zenodo.org/records/6008882} \\
    \midrule
Pragmatic MDR \cite{pragmaticMDR} & Emphasis of the data model: \textbf{representational }\\
              & Approach for metadata standardization: \textbf{bottom-up} \\
              & Year of journal publication: \textbf{2021} \\
              & Open-source code available: \textbf{no} \\ 
    \bottomrule
\end{tabularx}
\label{tbl.existing_iso11179_mdr}
\end{table}

\subsection{caDSR}\label{related.caDSR}
caDSR is an ISO 11179-based MDR developed by the National Cancer Institute in the US, part of the caCORE infrastructure for cancer informatics. caDSR defines a comprehensive set of standardized metadata descriptors and stores the common data elements (CDEs) for multiple cancer research projects \cite{caDSR}. The source code of caDSR is available on GitHub and under active maintenance by its developers, separated into its browser repository \footnote{\url{https://github.com/CBIIT/cadsr-cde-browser}} and its curator repository \footnote{\url{https://github.com/CBIIT/cadsr-cdecurate}}. The metadata model of caDSR contains both the conceptual and representational layers of the ISO 11179-3 core metadata model, rendering its metadata model more complete compared to Samply.MDR, QL$^4$MDR, and Pragmatic MDR. However, we see the following drawbacks in the CDE schema of caDSR:
\begin{itemize}
    \item Its metadata model differs from ISO 11179-3 in being less rigorous, leading to duplication and curation errors.
    \item Concepts lack hierarchy. Therefore, they exist at a single level.
    \item No inter-relationship between concepts which increases redundancy.
    \item The absence of a "synonyms/terms" table and the concepts are only classified by keywords, rendering searching less robust.
    \cite{nadkarni2006_caDSR_eval}
\end{itemize}

\subsection{Samply.MDR}\label{related.samplyMDR}

Samply.MDR \cite{SamplyMDR}, developed by the German Medical Informatics Initiative (MII) \cite{MII-DE}, provides its metadata repository publicly on GitHub (\url{https://github.com/samply}).
Although Samply.MDR aims for standardized, reusable data; it implements only the representational layer of the ISO 11179-3 model (shown in \cref{fig.iso11179_model}) with extended data types \textit{CalendarType}, \textit{NumericalType}, and \textit{StringType}, and additional classes \textit{Group}, \textit{Label}, and \textit{Namespace}. As Samply.MDR lacks a conceptual layer; this metadata repository lacks the findability of compatible features across data registries.

\subsection{QL4MDR}\label{related.ql4MDR}
QL$^4$MDR is an ISO 11179-based MDR designed as a technical interlingua for integrating heterogeneous EHR metadata \cite{ql4mdr}. 
The MDR uses a GraphQL-based query language and is only implemented at the representational layer.
Therefore, QL$^4$MDR has the same shortcomings as Samply.MDR, which makes the findability of compatible features across registries difficult.

\subsection{Pragmatic MDR}\label{related.pragmaticMDR}
Pragmatic MDR adopts a bottom-up standardisation approach that requires input from medical experts to ensure interoperability.
It harmonizes existing ISO 11179-based (caDSR and Samply.MDR) and non-ISO 11179-based (CancerGrid \cite{davies2014cancergrid}, CoMetaR \cite{stohr2017cometar}, and MDM Portal \cite{dugas2016_mdm_portal}) MDRs to create case report forms and routine documentation \cite{pragmaticMDR}.
Similar to Samply.MDR and QL$^4$MDR, Pragmatic MDR focuses on the representational layer of the ISO 11179-3 core metadata model and collects \textit{Data\_Elements} bottom-up. Pragmatic MDR provides 466,569 unique metadata definitions, including an API and a UI available to the public \footnote{\url{https://medical-data-models.org/MDR}}. However, its source code is not publicly available and lacks visualization of metadata definitions, limiting feature selection for ML studies.

\subsection{Drawbacks of existing MDRs}\label{related.drawback_summary}
To summarize, the current ISO 11179-based MDRs do not sufficiently improve the findability of trainable features in ML studies. Samply.MDR and caDSR display committee-approved top-down metadata, preventing bottom-up user updates. Meanwhile, QL$^4$MDR and Pragmatic MDR focus only on the representational layer, making it hard to assess compatibility between \textit{Data\_Element\_Concepts} across EHRs. No existing solutions for a middle-out approach that supports both representational and conceptual layers currently exist. To address these shortcomings, we developed a state-of-the-art ISO 11179-based MDR for the Screen4Care project \cite{S4C-EU}, called S4CMDR, to improve the findability of compatible and eventually co-trainable features across heterogeneous clinical data.
\section{Methods \& Implementation}\label{impl}
Due to the complex nature of medical ontology standards and EHR representations, we designed S4CMDR using the middle-out approach. This enables S4CMDR to utilize existing medical ontologies and support user-driven creation.
To minimize duplicate concepts derived from medical ontologies, we integrated the BioPortal API \cite{bioportal} into S4CMDR for automated suggestions of standardized vocabularies.
We modified the original ISO 11179-3 core metadata model to ensure compatibility and integration of BioPortal.
S4CMDR uses the microservice architecture to ensure scalability and flexibility.
The specific implementation details and design decisions are discussed in the following subsections.

\subsection{Integration with BioPortal} \label{impl.bioportal}
To support the middle-out approach, we require existing ontologies to suggest and minimize the number of duplicate concepts.
BioPortal, developed by the National Center for Biomedical Ontology, provides access to over 1,100 medical ontology standards via web and API.
It serves as an important reference to clinical MDRs mentioned in Section \ref{related.pragmaticMDR}.
BioPortal provides flexible and precise hierarchies in directed acyclic graphs (DAGs) found in medical ontology standards e.g., SNOMED CT, LOINC, ICD, and the NCI Thesaurus, ensuring that complex clinical relationships are captured accurately.

Integrating BioPortal into S4CMDR streamlines metadata creation and thereby avoids duplication.
S4CMDR matches the ontology class labels using BioPortal, therefore allowing users to automatically retrieve and annotate \emph{Conceptual\_Domains} and \emph{Data\_Element\_Concepts} efficiently. 
Class mappings provided by BioPortal are synonyms of each ontology class, allowing for further reduction of duplicates.

In addition, the DAG-like hierarchy of ontology classes used in BioPortal is crucial for maintaining semantic interoperability and the management of related medical ontology classes in S4CMDR.

\subsection{Modifications to the Core Metadata Model} \label{impl.core_data_model}
Despite the universal adoption of the ISO 11179-3 core metadata model for clinical MDRs, the practical limitations remain.
Ngouongo et al. \cite{Ngouongo_2013} found that flat registry structures and controlled clinical vocabularies fit the model well.
However, the semantic non-interoperability arose on complex EHR structures or classifications.
Therefore, the need for extending the ISO 11179-3 metadata model is necessary.

\begin{figure}[h] \centering \includegraphics[width=0.75\textwidth]{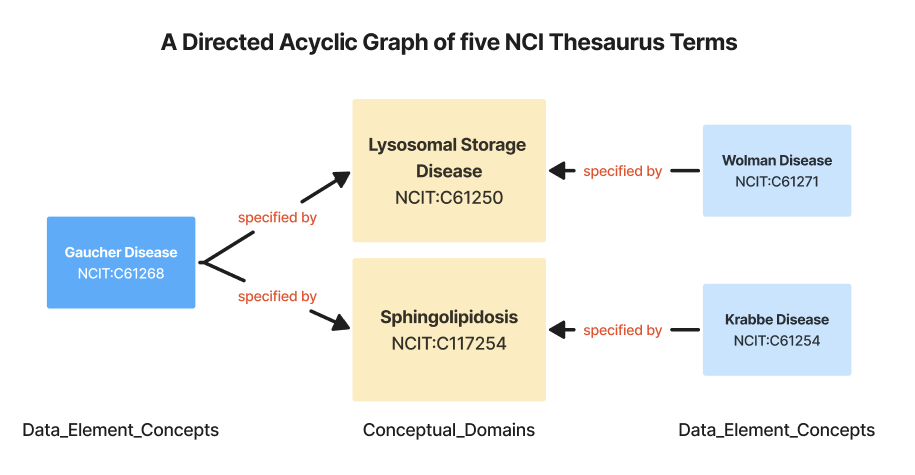}
\caption{This figure illustrates a partial directed acyclic graph (DAG) of three NCI Thesaurus terms that serve as \emph{Data\_Element\_Concepts} and two NCI Thesaurus terms that serve as \emph{Conceptual\_Domains}. The \emph{Data\_Element\_Concept} with more than one parent is shown in a blue rectangular block, the \emph{Data\_Element\_Concepts} each with only one parent are shown in light blue rectangular blocks, and \emph{Conceptual\_Domains} are shown in light yellow rectangular blocks.}
\label{fig.core_model_example_NCIT}
\end{figure}

\begin{figure}[h] \centering \includegraphics[width=0.75\textwidth]{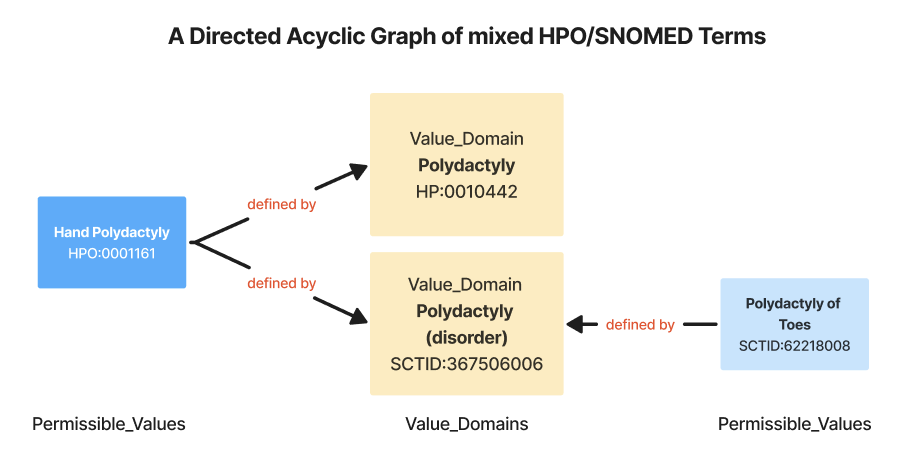}
\caption{This figure illustrates a partial directed acyclic graph (DAG) of two HPO or SNOMED CT terms that serve as \emph{Permissible\_Values} and two HPO or SNOMED CT terms that serve as \emph{Value\_Domains}. The \emph{Permissible\_Value} with more than one parent is shown in a blue rectangular block, the \emph{Permissible\_Value} with only one parent is shown in a light blue rectangular block, and \emph{Value\_Domains} are shown in light yellow rectangular blocks.}
\label{fig.core_model_example_HPO_SNOMED}
\end{figure}

\begin{figure}[h] \centering \includegraphics[width=0.9\textwidth]{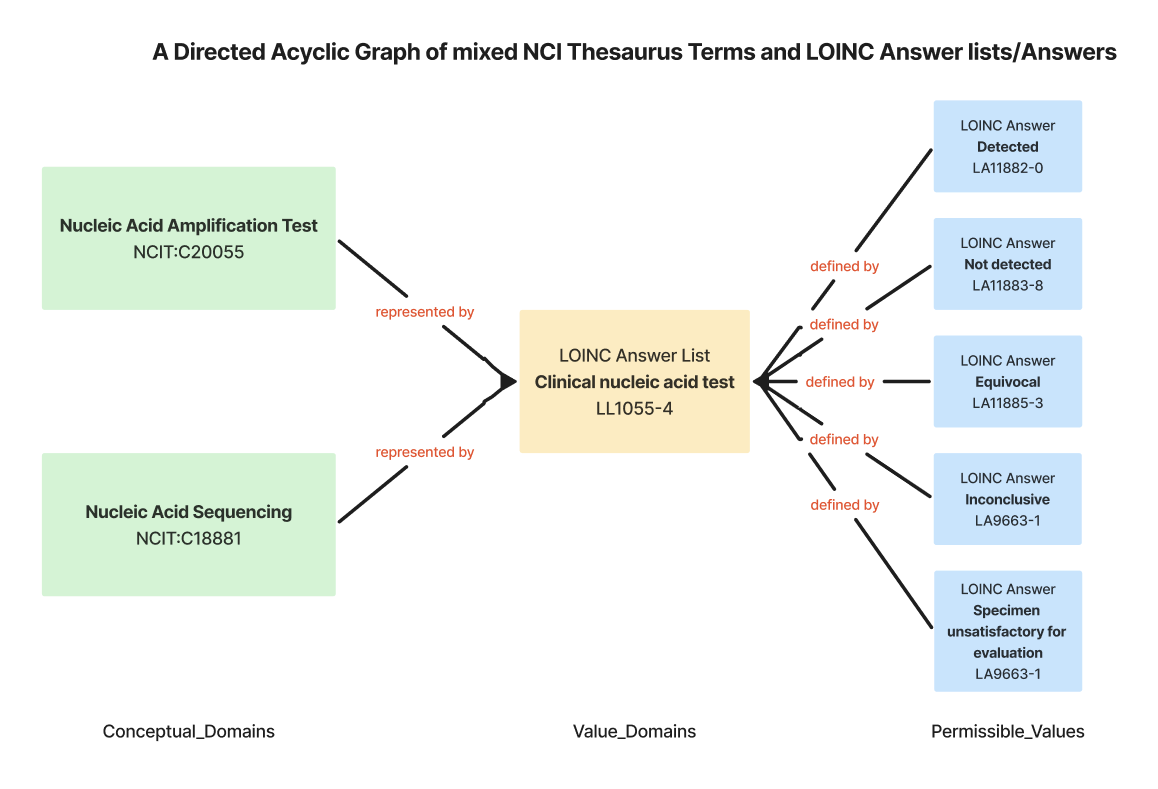}
\caption{This figure illustrates a partial directed acyclic graph (DAG) of two NCI Thesaurus terms that serve as \emph{Conceptual\_Domains}, one LOINC answer list that serves as a \emph{Value\_Domain}, and five valid LOINC answers that serve as \emph{Permissible\_Values}. The \emph{Conceptual\_Domains} are shown in light green rectangular blocks, the \emph{Value\_Domain} is shown in a light yellow rectangular block, and \emph{Permissible\_Values} are shown in light blue rectangular blocks.}
\label{fig.core_model_example_LOINC_NCIT}
\end{figure}

A major challenge is that various clinical ontologies, such as SNOMED CT \cite{Rodrigues2013_sharing_ontologies} and NCI Thesaurus \footnote{\url{https://ncit.nci.nih.gov/ncitbrowser/pages/multiple_search.jsf?nav_type=terminologies}}, follow DAGs \cite{snomedct_polyhierarchy, ncit_polyhierarchy} rather than trees as ISO 11179 intends.
We illustrate this with three concrete examples:
\begin{enumerate}
    \item A child term may have multiple parent terms. A terminal term corresponds to a \emph{Data\_Element\_Concept} while its parents correspond to \emph{Conceptual\_Domains}. For example, in the NCI Thesaurus, \emph{Gaucher's Disease} is a terminal term and is considered as \emph{Data\_Element\_Concept}, while its parents, \emph{Lysosomal Storage Disease} \footnote{\url{http://purl.obolibrary.org/obo/NCIT_C61250}} and \emph{Sphingolipidosis} \footnote{\url{http://purl.obolibrary.org/obo/NCIT_C117254}} serve as \emph{Conceptual\_Domains} (See \cref{fig.core_model_example_NCIT}). However, the original ISO 11179 disallows multiple parents (See \cref{fig.iso11179_model}).
    \item Human Phenotype Ontology (HPO) corresponds to \emph{Permissible\_Value} in ISO 11179-3, where most of the terms can be mapped to SNOMED CT \cite{Dhombres2016_snomed_hpo}. Consider HPO term \emph{Polydactyly} and SNOMED CT term \emph{Polydactyly (disorder)} as individual \emph{Value\_Domains}. It is possible to assign the \emph{Permissible\_Values} \emph{Polydactyly of Toes} and \emph{Hand Polydactyly} to the SNOMED CT term and reuse \emph{Hand Polydactyly} to assign it to the HPO term (See \cref{fig.core_model_example_HPO_SNOMED}). The current ISO 11179-3 does not allow mapping the same \emph{Permissible\_Values} to multiple \emph{Value\_Domains}.
    \item LOINC, a standard for laboratory codes, assigns each code an answer list of IDs. For example, the answer list code \emph{LL1055-4} contains multiple answers: for example, \emph{Detected} and \emph{Not detected}. The answer lists are considered a \emph{Value\_Domain} altogether named \emph{Clinical nucleic acid test}, and the answers \emph{Permissible\_Values} (See \cref{fig.core_model_example_LOINC_NCIT}). The \emph{Conceptual\_Domains} \emph{Nucleic Acid Amplification Test} and \emph{Nucleic Acid Sequencing} may use the same \emph{Value\_Domain}; however, the current ISO 11179-3 model does not support such a use case.
\end{enumerate}

Based on these examples, we have performed three modifications to the core data model of the ISO 11179-3 standard. 

Firstly, we have extended the originally defined \emph{one-to-many} bidirectional relationship between \emph{Conceptual\_Domain} and \emph{Data\_Element\_Concept} to a \emph{many-to-many} bidirectional relationship, illustrated in example (1) above. 

Secondly, motivated by example (2) above, we extended the originally defined \emph{one-to-many} bidirectional relationship between \emph{Value\_Domain} and \emph{Permissible\_Value} to a \emph{many-to-many} relationship to support the reusability of \emph{Permissible\_Value} in several \emph{Value\_Domain} without creating duplicates. An illustration based on example (2) of this modification is depicted in \cref{fig.core_model_example_HPO_SNOMED}.

Thirdly, we extended the \emph{one-to-many} bidirectional relationship between \emph{Conceptual\_Domain} and \emph{Value\_Domain} to a \emph{many-to-many} bidirectional relationship to facilitate DAG-like hierarchies of the common ontologies, responding to example (3) described previously. This example is illustrated in \cref{fig.core_model_example_LOINC_NCIT}.

\begin{figure}[h] \centering \includegraphics[width=0.9\textwidth]{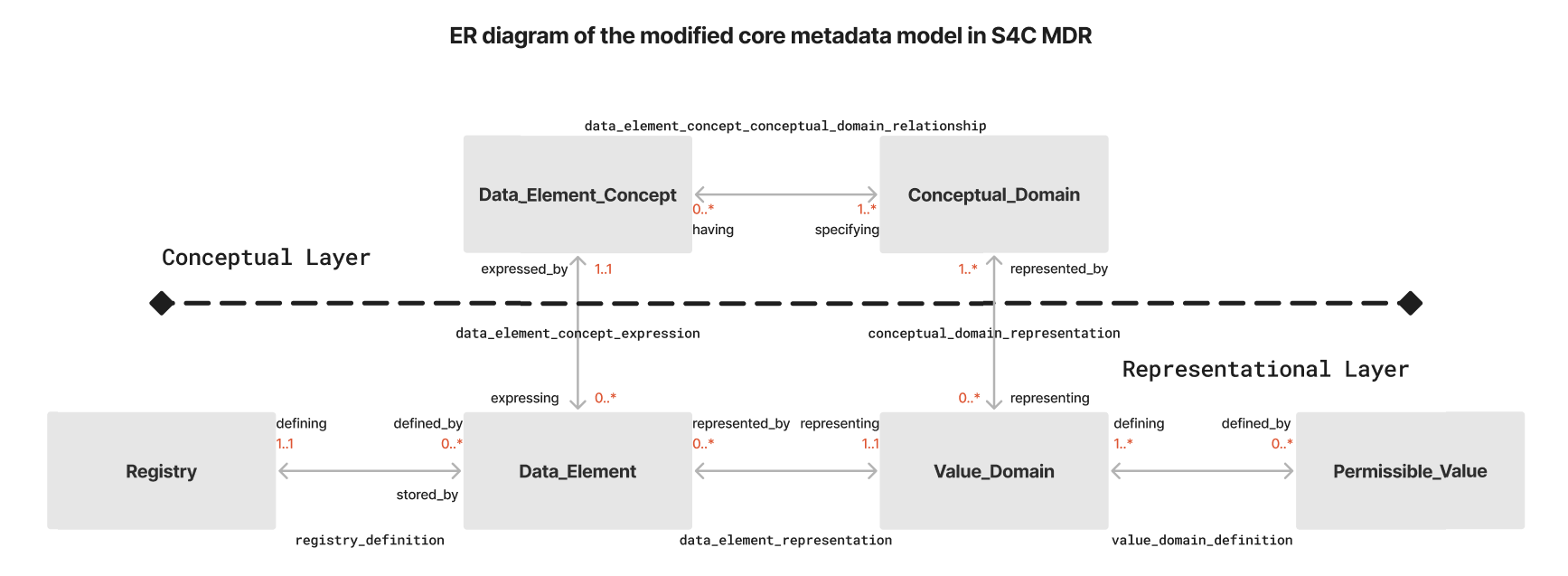} 
\caption{Compared to \cref{fig.iso11179_model}, three changes to the MDR core metadata model are made. Firstly, the mapping relation between \emph{Conceptual\_Domain} and \emph{Data\_Element\_Concept} is altered from \emph{one-to-many} to \emph{many-to-many}. Secondly, the mapping relation between \emph{Conceptual\_Domain} and \emph{Value\_Domain} is altered from \emph{one-to-many} to \emph{many-to-many}. Finally, the mapping relation between \emph{Value\_Domain} and \emph{Permissible\_Value} is altered from \emph{one-to-many} to \emph{many-to-many}.} 
\label{fig.s4c_mdr_data_model}
\end{figure}

\cref{fig.s4c_mdr_data_model} presents the general ER diagram of the modified core metadata model, which is based on the ISO 11179-3 standard and adapted for S4CMDR, including the mentioned key changes in the model.

\subsection{Software Architecture} \label{impl.microservices_arch}
\begin{figure}[h] 
    \centering 
    \includegraphics[width=0.9\textwidth]{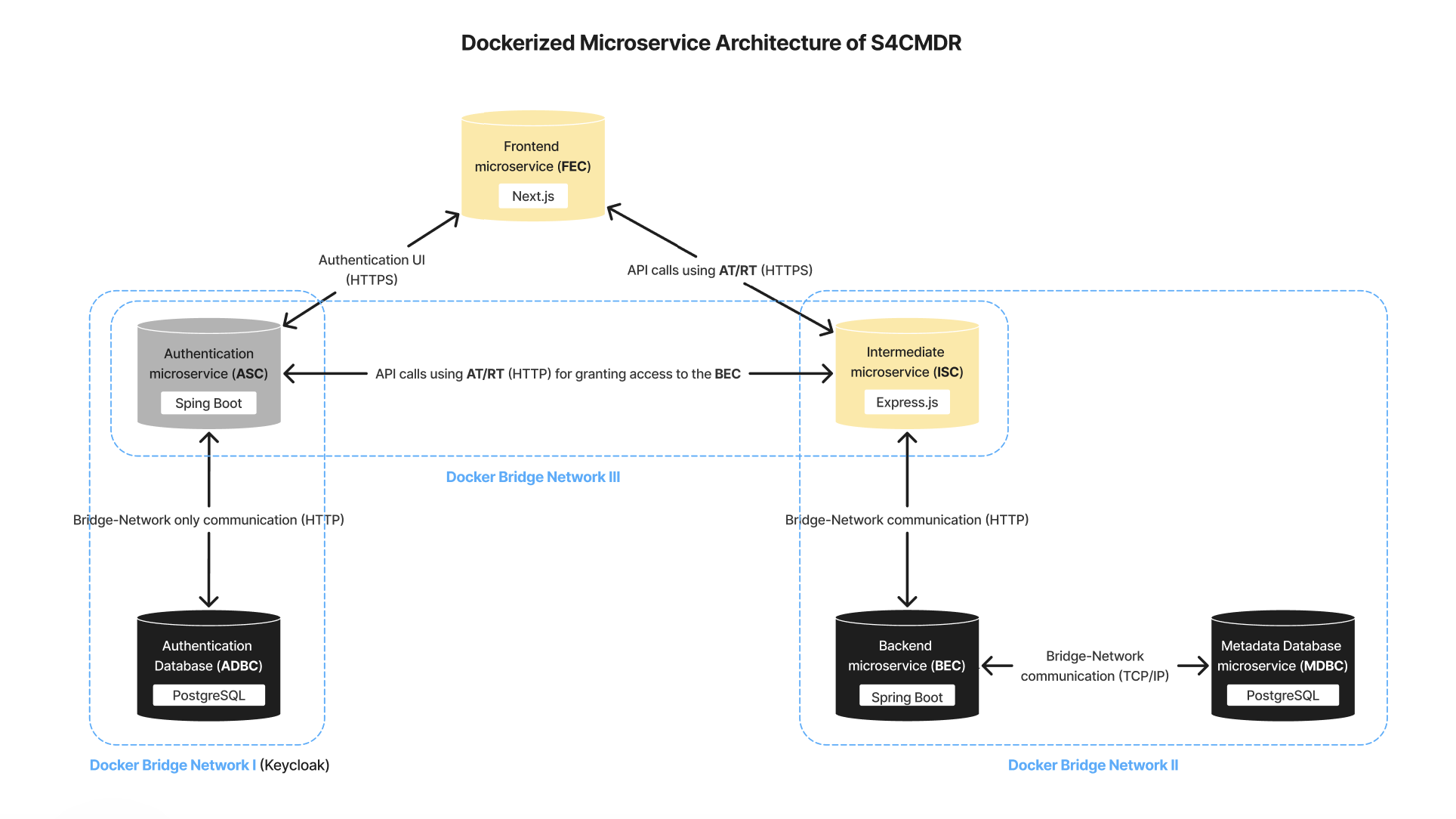} 
    \caption{The Docker containers highlighted in light yellow (FEC, ISC) are accessible for external HTTPS requests, the container colored in gray (ASC) provides only one URI accessible for the external HTTPS requests via the FEC (the sign-up/log-in user interface), and the containers colored in black are isolated from external access and can only be reached through the intermediate service container, providing an additional layer of security. Three separate Docker bridge networks are created to segregate different communication scenarios between the microservices, namely Docker containers and minimize the risk of malicious attacks. The Docker containers are deployed on our publicly available demonstration server on \url{https://s4cmdr.sdu.dk}.} 
    \label{fig.s4c_mdr_arch} 
\end{figure}

S4CMDR was developed following the latest best practices for web platform development and implemented as a flexible microservice architecture to ensure smooth maintenance and flexibility for our MDR. All microservices are containerized for easy deployment and maintenance, as shown in \cref{fig.s4c_mdr_arch}.

We carefully considered the technology stack for each container based on its use and functionality.
The microservice architecture consists of multiple containers, each with its own role:
\begin{itemize}
    \item The frontend container is built with Next.js, allowing users to interact directly with the S4CMDR.
    \item The container between the frontend and backend is referred to as the "intermediate service," which handles authentication checks, validates incoming requests for security, and fully integrates BioPortal for ontology support. This container is based on Express.js.
    \item The backend container based on SpringBoot \cite{arnold2005java} primarily concerns with retrieving and modifying data directly from the database.
    \item The database container is built to represent and support the modified ISO 11179-3 model using PostgreSQL.
\end{itemize}

In addition, we use Keycloak to enable authentication and user management.
Authorized users have access to certain metadata, while unauthorized users have restricted access.
Keycloak cooperates with the intermediate service to prohibit malicious users from accessing sensitive metadata.

\section{Results}\label{results}
Following the aforementioned considerations, we have implemented the S4CMDR. As a demonstration and validation, we started populating the database with publicly available data resources, with a particular focus on rare diseases. Our main objective is to support the researchers in conducting data-driven analyses (including e.g. machine learning), by removing roadblocks in the discovery of suitable and compatible data registries.

By merging the synonyms of an identical \emph{Data\_Element\_Concept} in different medical ontology standards, the number of redundant ontology items in the S4CMDR can be substantially reduced. Thus, the users can more accurately evaluate the number of features feasible for machine learning across data registries, which tends to be underestimated if not properly standardized and curated. We also encourage the users to populate the \emph{Value\_Domains} and \emph{Permissible\_Values} in S4CMDR with LOINC answer lists, and LOINC answers or HPO items, respectively. The multilingual coding system adopted by LOINC and HPO enables users to construct virtual patient cohorts for ML studies more effectively.

\subsection{Features of S4CMDR}
\subsubsection{Dashboard} \label{res.dashboard}
\begin{figure} \centering \includegraphics[width=0.9\textwidth]{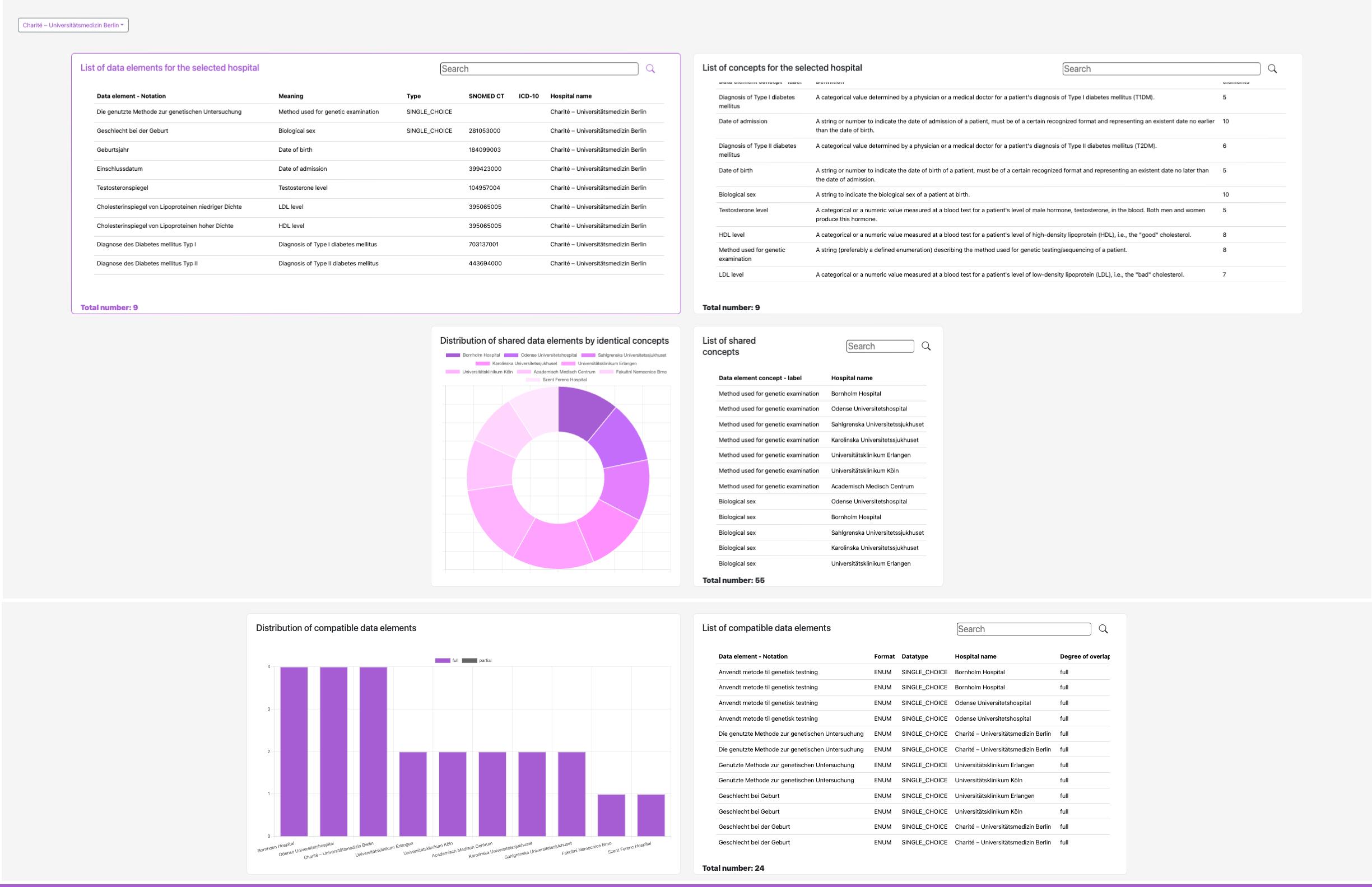} \caption{The S4CMDR dashboard, presenting a metadata summary through a combination of lists, a pie chart, and a bar chart, specifically tailored to the \emph{Registry} item "Charité – Universitätsmedizin Berlin."} \label{fig.ui_dashboard_charite} 
\end{figure}

After logging into S4CMDR, the users can view registry-specific summaries for metadata of electronic health records, such as the list of available \emph{Data\_Elements}, \emph{Data\_Element\_Concepts}, the number of shared \emph{Data\_Element\_Concepts} with every other data registry, and the number of "compatible" \emph{Data\_Elements} estimated by the \emph{Value\_Domains}. An example of the UI for registry-specific metadata summary on the frontend dashboard is shown in \cref{fig.ui_dashboard_charite}.

\subsubsection{CRUD operations} \label{res.CRUD}
Moreover, users with write privileges can perform create, read, update, and delete (CRUD) operations on metadata items defined in the S4CMDR core metadata model.

\subsubsection{Autosuggestions} \label{res.autosuggestions}
Autosuggestions are a crucial and helpful feature of S4CMDR, as they assist users by indicating semantically similar metadata items that already exist in S4CMDR or BioPortal. Users can thus be supported in minimizing the creation of redundant metadata items, thereby improving the findability of compatible features. 

\subsubsection{Minimisation of Duplicates} \label{res.miminize_duplicates}
Regarding the presentation of different medical ontologies in the database, when users enter data into the S4CMDR by selecting a relevant item suggested by the BioPortal API, we store its corresponding ontology name and ontology ID. We minimize the occurrence of duplicate entries in the database by enforcing a unique key constraint that combines the ontology name and ontology ID of each metadata item in the database schema.

\subsubsection{Compatibility function} \label{res.compatiblity_func}
With regard to the compatibility function in the S4CMDR, we compare similar \emph{Data\_Element} items that express the same \emph{Data\_Element\_Concept} items. Thus, we examine their corresponding \emph{Value\_Domain} items and \emph{Permissible\_Value} items. If their \emph{Value\_Domain} items match completely, we consider these \emph{Data\_Element} items to be fully compatible.
If the \emph{Data\_Element} items are associated with different \emph{Value\_Domain} items that share some of their \emph{Permissible\_Value} items, we consider these \emph{Data\_Element} to be partially compatible. In the case of partial compatibility, the backend of S4CMDR provides a method to create a temporary common \emph{Value\_Domain} item to achieve compatibility for training. Otherwise, we consider those \emph{Data\_Element} items to be fully incompatible.
\section{Discussions \& Conclusion}\label{discussion}
In this paper, we propose the S4CMDR, a modern metadata repository, developed to store and present the metadata of EHRs with a focus on medical ontologies. However, this is just the foundation to bring S4CMDR into real-world use for machine learning on EHRs, especially on EHRs of rare disease patients. Only with community support and a significant number of users who contribute to populating the MDR by inserting metadata or items from clinical dictionaries of real-world EHR-related datasets, the conspicuous evaluation of compatibility and visualisation across clinical data registries can be properly conducted. Otherwise, S4CMDR cannot be considered highly practical. Due to the fact that we have Screen4Care running as a large international research project, with 37 scientific, clinical and industry partners \footnote{\url{https://www.screen4care.eu/meet-the-partners/partners-of-screen4care}} that are potential users of the S4CMDR and populate it, external users will be able to leverage their substantial amount of existing clinical or scientific information, e.g., EHRs or other clinical studies, that can customize or tailor the S4CMDR to specific use cases in the academia and industry. Furthermore, we also invite other EHR holders in the EU or potentially outside the EU to populate S4CMDR using their registry-specific clinical data dictionaries.

A publicly available web frontend for S4CMDR on a server hosted by the University of Southern Denmark, a S4C project partner, at \url{https://screen4care.sdu.dk/}. Alternatively, the containerized API for authenticated users is available in \url{https://screen4care.sdu.dk/api/mdr}. We will provide continuous support and new features for the S4CMDR, e.g., user role management, more sophisticated options of CRUD operations, and front-end visualisation. In addition, we are currently working on the integration of the S4CMDR application with the Feasibility Query application for Screen4Care as containerized microservice to further facilitate the feature selection in ML and the construction of virtual patient cohorts for related researchers. We also invite researchers to register their own datasets within our S4CMDR in order to become findable for the rare disease community.

To conclude, S4CMDR is developed with a strong focus on providing an intuitive user interface and is designed following the idea of microservice architectures and containerisation principles, ensuring scalability and cross-platform compatibility. S4CMDR offers a friendly UI, registry-specific metadata visualisation, a functional API to metadata, configurable user authentication and authorisation management by Keycloak, and integration with the BioPortal API. The source code of S4CMDR is available under \url{https://gitlab.sdu.dk/screen4care/Metadata_repository}.

\section{Availability and requirements} \label{avail_and_req}
\textbf{Project name:} Screen4Care
\\
\textbf{Project home page:} \url{https://screen4care.eu}
\\
\textbf{Operating systems:} Linux
\\
\textbf{Programming languages:} Java, Typescript, SQL
\\ 
\textbf{Other requirements:} Docker
\\ 
\textbf{License:} GNU General Public License v3.0.
\\
\textbf{Any restrictions to use by non-academics:} no restrictions, please attribute work by citing this paper.

\section{Abbreviations} \label{abbre}
\textbf{\emph{CRUD:}} Create, Read, Update, and Delete
\\
\textbf{\emph{DAG:}} Directed Acyclic Graph
\\
\textbf{\emph{MDR:}} Metadata Repository
\\
\textbf{\emph{ML:}} Machine Learning
\\
\textbf{\emph{UI:}} User Interface

\begin{acks}
The Screen4Care EU-IMI project has received funding from the Innovative Medicines Initiative 2 Joint Undertaking (JU) under grant agreement No 101034427. The JU receives support from the European Union's Horizon 2020 research and innovation programme and EFPIA.
SBA Research (SBA-K1 NGC) is a COMET Center within the COMET – Competence Centers for Excellent Technologies Programme and funded by BMIMI, BMWET, and the federal state of Vienna. The COMET Programme is managed by FFG.
\end{acks}

\section{Author Contributions}
JZ, MA, NT, and RR worked on the design of the architecture and the core metadata model of the MDR. JZ, MA, and NT worked on the development of the S4CMDR application. JZ and RR wrote the manuscript. All authors read, corrected, and approved the final manuscript.
\bibliographystyle{ACM-Reference-Format}
\bibliography{sn-ref}

\end{document}